\documentclass[twocolumn,preprintnumbers,amsmath,amssymb]{revtex4}

\usepackage{graphicx}% Include figure files
\usepackage{dcolumn}% Align table columns on decimal point
\usepackage{bm}% bold math
\begin{document}
\newcommand{\pr}{\partial}
\newcommand{\ep}{\epsilon}
\newcommand{\p}{\prime}
\newcommand{\w}{\omega}
\newcommand{\lt}{\mathcal{L}}
\newcommand{\Fa}{F\left[\begin{array}{c|c} x & x^\p \\ t+\Delta t & t \end{array} \right]}
\newcommand{\Fb}{F\left[\begin{array}{c|c} x_t & x_0 \\ t & t_0 \end{array} \right]}

%------------------------------------------------------------------------------
\title{Onsager-Machlup theory and work fluctuation theorem for a harmonically driven    Brownian particle.}
\author{Navinder Singh}
\email{navinder@iopb.res.in}
\affiliation{Institute of physics, Bhubaneswar-751005, India} 

\begin{abstract}
We extend Tooru-Cohen analysis for nonequilirium steady state(NSS) of a Brownian particle to nonequilibrium oscillatory state (NOS) of Brownian particle by considering time dependent external drive protocol. We consider an unbounded charged Brownian particle in the presence of oscillating electric field  and prove work fluctuation theorem, which is valid for any initial distribution and at all times. For harmonically bounded and constantly dragged Brownian particle considered by Tooru and Cohen, work fluctuation theorem is valid for any initial condition(also NSS), but only in large time limit. We use Onsager-Machlup Lagrangian with a constraint to obtain frequency dependent work distribution function, and describe entropy production rate and properties of dissipation functions for the present system using Onsager-Machlup functional.
\end{abstract}  
\maketitle
\section{Introduction}
The first theory of fluctuations in equilibrium thermodynamics was given by A. Einstein, who used the Boltzmann relationship between the probability of a state and it's entropy. It states that the probability of a fluctuation from equilibrium state is proportional to $e^{\Delta S/k_B}$, where $\Delta S $ is the variation of entropy calculated along a reversible transformation creating the fluctuation and $k_B$ is the Boltzmann constant. 

In 1953, L. Onsager and S. Machlup generalized the equilibrium Boltzmann-Einstein formula to near equilibrium dynamical setting (near equilibrium irreversible processes) by using a variational principle to calculate the most probable trajectory followed by the system in the spontaneous emergence of a fluctuation\cite{onsager1,onsager2,casimir3,om1,om2}. The laws of correlations in average course of a fluctuation (a sequence of statistically correlated nonequilibrium states) are precisely given by the laws of irreversible processes. Under the assumption of time reversibility of microscopic dynamics, they showed that most probable creation and relaxation trajectories of a fluctuation are time reversal of one another. So, the path integral concept or the Onsager-Machlup functional gives a natural generalization to the time domain of Einstein-Boltzmann formula.

Recently, fluctuation theorems has drawn considerable attention in nonequilibrium statistical mechanics. Fluctuation theorems are the statements about the asymmetry of the distribution functions of work, heat etc. around zero, i.e., they involve negative tails of work or heat. These relations points towards rare events in macroscopic systems, which are more easily observable in small systems with finite degrees of freedom. These theorems give more complete picture of the second law of thermodynamics\cite{dje93,dje94,galco95,kur98,lebo99,crooks99,crooks00,cil98,wang02,cil04,feitosa04,gar05,schuler05,gallavotti96,gallavotti97,jar97,van03,van04,sek98,cohen03,cohen04,dhar04,maes,seif05,wij05}. It is known that near equilibrium fluctuation-dissipation relation and Onsager reciprocal relations can be obtained from fluctuation theorems\cite{lebo99,gallavotti96}.

In a recent work\cite{tooru} Tooru Taniguchi and E. G. D. Cohen extended Onsager-Machlup theory of fluctuations around equilibrium to a theory of fluctuations around nonequilibrium steady state for a specific system, consisting of a constantly dragged Brownian particle (B-particle) through fluid. They discuss thermodynamics of this system using generalized Onsager-Machlup Lagrangian. They also discuss fluctuation theorems based on their generalized Onsager-Machlup theory and generalized forms of the detailed balance conditions for nonequilibrium steady states. A very general treatment of macroscopic fluctuation theory for a class of continuous time Markov chains is given by  Bertini etal\cite{berti02,berti06}, and a general treatment considering large deviation theories is given by B. Derrida etal\cite{derrida02,derrida07}.
% write about Derrida also.

The present paper has a limited objective, in this work we extend Tooru-Cohen analysis for nonequilirium steady state(NSS) of B-particle to nonequilibrium oscillatory state (NOS) of B-particle by considering time dependent external drive protocol. We consider an unbounded charged Brownian particle in the presence of oscillating electric field (figure 1) and prove work fluctuation theorem, which is valid for any initial distribution and at all times. We use path integral approach similar to Tooru-Cohen approach\cite{tooru}. We use Onsager-Machlup Lagrangian with a constraint to obtain frequency dependent work distribution function.

The paper is organized as follows. In section II we introduce our model and finite time transition probability using path integral approach. We then calculate it's most probable path using variational principle and solve functional integral for finite time transition probability. In section III we calculate work distribution function and prove work fluctuation theorem. In section IV we use Onsager-Machlup functional to describe coarsed-grained B-particle thermodynamics. In section V we end with conclusion.
%------------------------------------fig.No.1 --------------------------
\begin{figure}
\includegraphics[height=5cm, width=5cm]{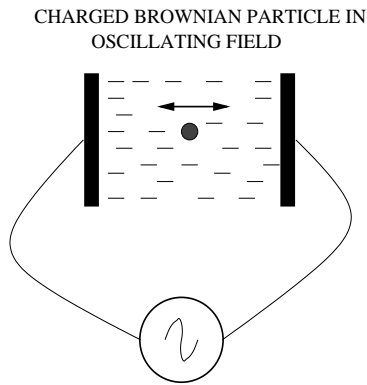}
\caption{Charged Brownian particle in an oscillating electric field}
\end{figure}
%----------------------------------------------------------------------
\section{A charged Brownian particle in an oscillating electric field}
Consider a charged Brownian particle in an oscillating electric field as shown in figure 1. We assume that the external harmonic drive time scale is much larger than internal dynamical time scale( inverse molecular collision frequency) of the B-particle-fluid system, i.e.,
\begin{equation}
\frac{1}{\omega} \gg \frac{1}{molecular~collision~frequency}.
\end{equation}
Thus friction coefficient is independent from drive frequency\cite{lebo65}. The B-particle's driven  stochastic dynamics is given by the following Langevin equation,
\begin{equation}
m\frac{d^2x_t}{dt^2} = -\alpha\frac{dx_t}{dt} -qE_0 cos\omega t + \xi_t
\end{equation}
Here, $\xi_t$ is a Gaussian-White Noise(GWN) with properties $\langle \xi_t\rangle=0$ and $\langle \xi_t \xi_{t^\p}\rangle=2\alpha/\beta\delta(t-t^\p) $.  In an overdamped case, it is
\begin{equation}
\frac{dx_t}{dt}=-\frac{qE_0}{\alpha}cos\omega t +\frac{1}{\alpha}\xi_t.
\end{equation}
%-------------------------------
The Fokker-Planck Equation for the above Langevin equation is[see Appendix A for derivation]
\begin{equation}
 \frac{\pr f}{\pr t} = \frac{\pr}{\pr x} \eta\cos\omega t f + D \frac{\pr^2 f}{\pr x^2}=\hat{\lt} f.
\end{equation}
Here $f$ is the noise averaged distribution function, $D=\frac{1}{\alpha\beta}$ is the diffusion coefficient, with $\beta = 1/{k_B T}$, and $\eta =\frac{qE_0}{\alpha}=\frac{F_e}{\alpha}$.
The Fokker-Planck operator $\hat{\lt} $ is defined as
\begin{equation}
\hat{\lt} = \frac{\pr}{\pr x} \eta\cos\omega t + D \frac{\pr^2}{\pr x^2}.
\end{equation}
%\subsection{Transition probability}
The transition probability for B-particle for a small step from $(x^{\p}, t)$ to $(x, t+\Delta t)$ is given by
\begin{equation}
\Fa = e^{\hat{\lt} \Delta t}\delta(x-x^\p)
\end{equation}
Transition probability for a finite time interval[see Appendix B for a complete calculation] is given by
\begin{equation}
\Fb =\int_{x_0}^{x_t}Dx_s\exp{\left[\int_{t_0}^t ds L(\dot{x}_s) \right]}.
\end{equation}
The Lagrangian $L$ for this driven Brownian motion is;
\begin{equation}
 L(\dot{x}_s)=-\frac{1}{4D}(\dot{x}_s +\eta\cos\omega s)^2,
\end{equation}
and the Path Integral is introduced  as;
\begin{eqnarray}
&&\int_{x_0}^{x_t}Dx_s\exp{\left[\int_{t_0}^t ds L(\dot{x}_s) \right]}= \lim_{N\rightarrow +\infty}\left(\frac{1}{\sqrt{4\pi D \Delta t}}\right)^N\nonumber\\
&&\times\int dx_{t_{N-1}}\int dx_{t_{N-2}}\ldots\int dx_{t_{1}}\exp{\left[\int_{t_0}^t ds L(\dot{x}_s)\right]}.\nonumber\\
\end{eqnarray}
Here the initial time is $t_0$, the final time is $t_N = t$, and the initial position  is $x_0$, the final position is $x_t$, with $t_n = t_0 + n\Delta t,\,\, n = 1,2,3,...,N,\,\,\,\Delta t = (t-t_0)/N$.
%-----------------------------------------
\subsection{Solution Path: Calculation of the above Functional Integral}
In this stochastic dynamical setting where the Brownian particle is  fluctuating about the most probable path governed by external harmonic drive, the variational properties of the action governs the most probable trajectory of the particle between two given fixed points of the path. In the present setting the variational principle takes care of two aspects of the B-particle motion, (1) the systematic part due to harmonic drive and (2) the stochastic part due to molecular impacts(which further involves fluctuation and dissipation). Is is to be noted here that the dissipative aspect of the motion which originates from two different physical effects (1) due to stochastic noise(fluctuation-dissipation), and (2) due to external systematic drive, are taken into account by phenomenological parameter $\alpha$ in the Langevin equation, and in the present treatment we are not taking into account the frictional force which is nonlinear in velocity possible for an arbitrary drive strength. The variation principle which maximizes the transition probability i.e., for the solution path($\{\tilde{x}_s\}$, from $x_0$ to $x_t$) is
\begin{equation}
 \delta\int_{t_0}^{t} ds L(\dot{x}_s) = 0,
\end{equation}
which implies Euler-Lagrange equation
\begin{equation}
\frac{d}{ds}\left(\frac{\pr L}{\pr\dot{\tilde{x}}_s}\right) -\frac{\pr L}{\pr\tilde{x}_s} = 0.
\end{equation}
The solution path obtained from above the equation with end points at $\tilde{x}_0 = x_0$ and at $\tilde{x}_t = x_t$ is given by,
\begin{equation}
\tilde{x}_s = -\frac{\eta}{\omega}\sin\omega s + \left(\frac{x_t +\frac{\eta}{\omega}\sin\omega t -x_0}{t}\right) s +x_0.
\end{equation}
To calculate the functional integral in equation (9), we will first calculate the integral $\left[\int_{t_0}^t ds L(\dot{x}_s) \right]$. For this, consider a variation $\tilde{z}_s \equiv x_s-\tilde{x}_s$ about the solution path ${\tilde{x}_s}$
\begin{equation}
 \int_{t_0}^{t} ds L(\dot{\tilde{x}}_s+\dot{\tilde{z}}_s)=\int_{t_0}^{t} ds[L+\dot{\tilde{z}}_s\frac{\pr L}{\pr \dot{\tilde{x}}_s}+\frac{1}{2}(\dot{\tilde{z}}_s)^2\frac{\pr^2 L}{\pr{\tilde{x}_s}^2}+\ldots]
\end{equation}
Inserting various derivatives from the known Lagrangian and solution path, the above integral is
\begin{equation}
=\int_{t_0}^{t} ds L(\dot{\tilde{x}}_s) -\frac{1}{4D}\int_{t_0}^{t}ds[2(x_t+\frac{\eta}{\omega}\sin\omega t-x_0)+\dot{\tilde{z}}_s]\dot{\tilde{z}}_s.
\end{equation}
The functional integral (equation 9) then becomes
\begin{eqnarray}
&&\Fb = e^{-\frac{1}{4Dt}g(t)^2}\nonumber\\
&&\times\int_{\tilde z_0}^{\tilde z_t} D{\tilde{z}_s}\exp{\left[-\frac{1}{4 D}\int_{t_0}^t ds (2 g(t)/t + \dot{\tilde{z}}_s)\dot{\tilde{z}}_s\right]}.
\end{eqnarray}
Here $g(t)=x_t+\frac{\eta}{\omega}\sin\omega t-x_0$. Next we calculate the functional integral in the above equation(Eq.15) for which $\tilde z_0 (t_0 = 0) =\tilde z_t = 0$. Writing the exponential part as
\begin{eqnarray}
&&\exp{\left[-\frac{1}{4 D}\int_{t_0}^t ds
\left((\dot{\tilde{z}}_s+g(t)/t)^2-(g(t)/t)^2\right)\right]}\nonumber\\
&&\simeq \exp{\left[-\frac{1}{4 D\Delta t}\sum_{i = 1}^N 
\left(\tilde{z}_{i+1}-\tilde{z}_{i}+\Delta t g(t)/t)\right)^2\right]}e^{\frac{g(t)^2}{4 D t}}\nonumber\\.
\end{eqnarray}
\begin{equation}
=e^{\left[-\frac{1}{4 D\Delta t}\sum_{i = 1}^N 
(\tilde{z}_{i+1}-\tilde{z}_{i})^2-\frac{1}{4 D}(N+1)\Delta t (\frac{g(t)}{t})^2\right]}e^{\frac{g(t)^2}{4 D t}}.
\end{equation}
On taking the limit $\Delta t \rightarrow 0$ or $N \rightarrow \infty$ with $t-t_0 = N\Delta t$, the first part of the exponential after path integration gives standard result $1/\sqrt{4\pi D t}$(for standard solution see\cite{wie}), and the second gives $e^{-\frac{g(t)^2}{4 D t}}$, we finally obtain
\begin{equation}
 \Fb = \frac{1}{\sqrt{4\pi D t}}e^{-\frac{1}{4Dt}(x_t-x_0+\frac{\eta}{\omega}\sin\omega t)^2}.
\end{equation}
Which reduces to free B-particle motion for $\eta = 0 (no ~ external ~drive)$.
%%%%%%%%%%%%%%%%%%%%%%%%%%%%%%%%%%%%%%%%%%%%%%%%%%%%%%
\section{Functional integral calculation for  work distribution function}
The rate of work done by the external oscillating electric field is given by
\begin{equation}
\dot{W}(\dot{x}_t)=(-qE_0\cos\omega t)\dot{x_t}.
\end{equation}
For simplicity we consider the dimensionless work $\beta W_t(\{x_s\})$ and it's distribution defined as 
\begin{equation}
P_W(W,t) \equiv \langle\langle\delta(W - \beta W_t(\{x_s\}) )\rangle\rangle_t. 
\end{equation}
Here, $\langle\langle\ldots\rangle\rangle_t$ means a functional average over all possible paths $\{x_s\}$ between starting and end points as well as integrals over all starting and end points of the path:
\begin{eqnarray}
&&P_W(W,t) = \int dx_t\int_{x_0}^{x_t} Dx_s\int dx_0 f(x_0,t_0) e^{\int_0^t ds L(\dot{x}_s)}\nonumber\\
&&\times [\delta(W - \beta W_t(\{x_s\}) )].
\end{eqnarray}
The term $e^{\int_0^t ds L(\dot{x}_s)}$ is the probability functional for the path  which emanate from any initial distribution denoted as $f(x_0,t_0)$. Inserting integral representation for delta function in the above equation (Eq. 21) we get
\begin{eqnarray}
&&P_W(W,t)=\frac{1}{2\pi}\int_{-\infty}^{+\infty}d\lambda\int dx_t\int_{x_0}^{x_t} Dx_s\int dx_0 \nonumber\\
&& \times f(x_0,t_0) e^{\int_0^t ds L(\dot{x}_s)}e^{i\lambda W - i\lambda\beta W_t(\{x_s\})}
\end{eqnarray}
Inserting the expression $W_t(\{x_s\})=\int_0^t ds \dot{W}(\dot{x}_s)$ for work done in a finite time interval in the above equation, we obtain
\begin{eqnarray}
&&P_W(W,t)=\nonumber\\
&&\frac{1}{2\pi}\int_{-\infty}^{+\infty}d\lambda e^{i\lambda W}\int dx_t\int dx_0 f(x_0,t_0) F(x_t,x_0;i\lambda).\nonumber\\
\end{eqnarray}
Where $F(x_t,x_0;\lambda)$ is the constrained transition probability
\begin{equation}
 F(x_t,x_0;\lambda)\equiv\int_{x_0}^{x_t}Dx_s\exp{\left\{\int_0^t ds [L(\dot{x}_s)-\lambda\beta\dot{W}(\dot{x}_s)]  \right\}}
\end{equation}
With, $L(\dot{x}_s)-\lambda\beta\dot{W}(\dot{x}_s)\equiv L_{total}$. Next we calculate the solution path for the total Lagrangian from Euler-Lagrange equation, and then calculate the constrained transition probability $F(x_t,x_0;\lambda)$. Solving the Euler-Lagrange equation (Eq. 11) for the above total Lagrangian we obtain the solution path as
\begin{equation}
\tilde{x}_s = -\frac{\eta(1-2\lambda)}{\omega}\sin\omega s + \left(\frac{x_t +\frac{\eta(1-2\lambda)}{\omega}\sin\omega t -x_0}{t}\right) s +x_0.
\end{equation}
Proceeding on exactly similar lines for the calculation of transition probability (from equation 13 to 18) we obtain constrained (taking work rate into account) transition probability as
\begin{eqnarray}
&&F(x_t,x_0;\lambda)=\nonumber\\
&&\frac{1}{\sqrt{4\pi Dt}}\exp{\left\{-\frac{1}{4 D t}(g_1(t,\lambda))^2 -\lambda g_2(t)+\lambda^2 g_2(t) \right\}}\nonumber\\
&& g_1(t,\lambda)=x_t-x_0+(1-2\lambda)\frac{\eta}{\omega}\sin\omega t\nonumber\\
&& g_2(t) = \frac{\eta^2}{2 D}(t+\frac{\sin 2\omega t}{2\omega}).
\end{eqnarray}
Which reduces to the previous result (equation 18) for $\lambda = 0$. The calculation for $P_W(W,t)$ is straightforward, inserting equation 26 into equation 23, and using standard Gaussian integrals and normalization condition for initial distribution, we have
\begin{equation}
P_w(W,t)= \frac{1}{\sqrt{8\pi \frac{\eta^2}{2 D}(1+\frac{\sin 2\omega t}{2\omega t})t}}e^{-\frac{\left(W-\frac{\eta^2}{2 D}(1+\frac{\sin 2\omega t}{2\omega t})t\right)^2}{4 \left[\frac{\eta^2}{2 D}(1+\frac{\sin 2\omega t}{2\omega t})t\right]}}.
\end{equation}
From which we immediately obtain work fluctuation theorem:
\begin{equation}
 \frac{P_W(W,t)}{P_W(-W,t)} = e^W.
\end{equation}
The important point is that we have proved the work fluctuation theorem for any initial distribution and for all times as an identity (for a free driven B-particle, where the electrostatic force acts directly on the particle\cite{com1}). For the system (a harmonically bounded and constantly dragged Brownian particle) considered by Tooru and Cohen\cite{tooru}, work fluctuation theorem is valid for any initial condition(also NSS), but only in large time limit.
%\section{Harmonically bounded charged Brownian particle in oscillating electric field}
%\subsection{Extended heat fluctuation theorem}
\section{Onsager-Machlup functional and Coarse-Grained Brownian Particle Thermodynamics}
From equations (7) and (8), the Onsager-Machlup (OM) Lagrangian function is defined as
\begin{equation}
L(\dot{x}_s)= -\frac{\alpha}{4 k_B T}\left[\dot{x}_s + \frac{F_e}{\alpha} \cos \omega s\right]^2,
\end{equation}
where $F_e = q E_0$. The above OM Lagrangian can be expressed as
\begin{equation}
L(\dot{x}_s)= -\frac{1}{2 k_B} \left[\Phi(\dot{x}_s)+\Psi(F_e,\omega)-\dot{S}(\dot{x}_s)\right],
\end{equation}
where $\dot{S}(\dot{x}_s)\equiv-\frac{1}{T}(\dot{x}_s)F_e\cos\omega s$ is the Entropy production rate, $\Phi(\dot{x}_s)\equiv\frac{\alpha}{2 T}(\dot{x}_s)^2$ and $\Psi(F_e,\omega)\equiv\frac{1}{2T}\frac{F_e^2}{\alpha}\cos^2\omega s$ are Dissipation functions. The entropy production rate $\dot{S}(\dot{x}_s)$ is a positive function for a coarsed grained description
\begin{equation}
 Coarse-graning\equiv\langle\ldots \rangle_{noise}.
\end{equation}
Using Langevin equation (Eq. (3)) we obtain
\begin{equation}
 \dot{S}(\langle\dot{x}\rangle_{noise})=\frac{F_e^2}{\alpha T}\cos^2\omega t\geq 0,
\end{equation}
which justifies the property of entropy production rate. It is clear from the definition of Dissipative Functions $\Phi(\dot{x}_s)\equiv\frac{\alpha}{2
T}(\dot{x}_s)^2$ and    $\Psi(F_e,\omega)\equiv\frac{1}{2T}\frac{F_e^2}{\alpha}\cos^2\omega s$ that they are invariant under the time reversal change $\dot{x}_s \rightarrow -\dot{x}_s$, and both are positive functions. For noise averaged velocity (from noise averaging of Langevin equation) both the dissipation functions are equal to each other,
\begin{equation}
 \Psi(\langle\dot{x}_s\rangle_{noise})=\Phi(F_e,\omega)=\frac{1}{2T}\frac{F_e^2}{\alpha}\cos^2\omega s \geq 0.
\end{equation}
From Eq. (32) and Eq. (33) we obtain
\begin{equation}
\dot{S}(\langle\dot{x}\rangle_{noise})= 2\Phi(\langle\dot{x}_s\rangle_{noise})=2\Psi(F_e,\omega) \geq 0.
\end{equation}
In other words the sum of dissipation functions is equal to the rate of total entropy production.

Next we will show that the energy conservation law is true for any fluctuation(without coarse-graning). The rate of work done on the B-particle by the external field is defined as $\dot{W}(\dot{x}_s)\equiv -F_e (\dot{x}_s)\cos\omega s$, and the total work done for a time interval $t$ is given by $W_t(\{x_s\})=\int_0^t ds \dot{W}(\dot{x}_s)$. The heat generated $Q_t(\{x_s\})$ for the solution path $\{x_s\}$ is defined as
\begin{eqnarray}
&&Q_t(\{x_s\})\equiv T\int_0^t ds \dot{S}(\dot{x}_s)=-\int_0^t ds \dot{x}_s F_e\cos\omega s\nonumber\\
&& = \int_0^t ds \dot{W}(\dot{x}_s)=W_t(\{x_s\}).
\end{eqnarray}
Thus for a steady oscillatory state, the work done by the external field is dissipated as heat into the fluid surrounding the Brownian particle.
%\subsection{Free B-particle case}
%\subsection{Bounded B-particle case}
%\section{Effect of inertia}
\section{Conclusion}
For an unbounded charged Brownian particle in the presence of oscillating electric field we have proved work fluctuation theorem, which is valid for any initial distribution and at all times. We have used path integral approach similar to Tooru-Cohen approach and extend their analysis for nonequilirium steady state(NSS) of  a Brownian particle to nonequilibrium oscillatory state (NOS) of Brownian particle by considering time dependent external drive protocol. For harmonically bounded and constantly dragged Brownian particle considered by Tooru and Cohen, work fluctuation theorem is valid for any initial condition, but only in large time limit. We used Onsager-Machlup Lagrangian with a constraint to obtain frequency dependent work distribution function, and used Onsager-Muchlup functional to describe thermodynamic properties of the present system. A complete treatment of this problem valid at any distance from equilibrium(when frictional force is nonlinear in velocity) and at all frequencies\cite{lebo65} is an open issue.
\section{Acknowledgment}
I am tankful to Nabyendu Das for helpful remarks.
%%%%%%%%%%%%%%%%%%%%%%%%
\section*{Appendix A: Fokker-Planck Equation}
To drive the Fokker-Planck Equation\cite{zwan} for the  Langevin equation(Eq.(3)), we  define probability distribution of variable $x$  at time $t$ as $f(x,t)$. Probability conservation gives $\int dx f(x,t) = 1$, or
\begin{equation}
\frac{\pr f}{\pr t} + \frac{\pr}{\pr x}(\frac{\pr x}{\pr t} f)=0 
\end{equation}
\begin{equation}
\frac{\pr f}{\pr t}=-\frac{\pr}{\pr x}(-\eta \cos \omega t f +\frac{1}{\alpha} \xi_t f)
\end{equation}
Define $\hat{L} f = \frac{\pr}{\pr x}(-\eta \cos \omega t) f,~ ~ with~ \eta = \frac{q E_0}{\alpha}$. First we will consider the noise free part of the above equation i.e., $\frac{\pr f}{\pr t} = -\hat{L} f$, it's solution is
\begin{equation}
 f(x,t) = e^{-\int_0^t ds \hat{L}} f(x,0).
\end{equation}
With noise term,
\begin{equation}
\frac{\pr f}{\pr t} = -\hat{L} f -\frac{\pr}{\pr x}\frac{1}{\alpha}\xi_t f,
\end{equation}
we have a solution of the form
\begin{eqnarray}
&&f(x,t)=e^{-\int_0^t dt^\p \hat{L}}f(x,0)-\int_0^t ds e^{-\int_0^t dt^\p \hat{L}} e^{\int_0^s dt^{\p\p} \hat{L}} \nonumber\\
&&\times\frac{\pr}{\pr x}\frac{1}{\alpha}\xi_s f(x,s)
\end{eqnarray}
Substituting Eq. (40) back into Eq. (39), and taking average over noise, we obtain
\begin{eqnarray}
&&\frac{\pr \langle f(x,t)\rangle}{\pr t} = -\hat{L} \langle f(x,t)\rangle \nonumber\\
&&+ \frac{1}{\alpha^2}\frac{\pr}{\pr x}\int_0^t ds e^{-[\int_0^t dt^\p \hat{L}-\int_0^s dt^{\p\p}\hat{L}]}\langle \xi_t\xi_s\rangle\frac{\pr}{\pr x}\langle f(x,s)\rangle\nonumber\\
\end{eqnarray}
It is important to note that $f(x,t)$ depends on noise $\xi_s$ only for times $s$ that are earlier that $t$. The RHS of the above equation contains two explicit noise factors, $\xi_t$ and $\xi_s$, and also those {\em earlier time} noise factors that are implicit in $f(x,s)$. The first pair gives $\delta(t-s)$, and second  pair with the implicit noise factor gives $\delta(t-s^{\p})$, which does not contribute to the above integral as $t>s>s^\p$. Only first pair will contribute. Using properties of Gaussian white noise and delta function, we obtain
\begin{equation}
 \frac{\pr f}{\pr t} = \frac{\pr}{\pr x}\eta\cos\omega t f + D \frac{\pr^2 f}{\pr x^2}.
\end{equation}
Here $f$ is the noise averaged distribution function and diffusion coefficient $D = 1/{\alpha\beta}$, with $\beta = 1/{k_B T}$.
\section*{Appendix B: Transition Probability}
The transition probability for B-particle for ``a small time step'' $\Delta t$ from $(x^\p, t)$ to $(x, t+\Delta t)$ is given by\cite{ris}
\begin{equation}
\Fa = e^{\lt \Delta t}\delta(x-x^\p)
\end{equation}
\begin{eqnarray}
&& \Fa =\exp\left[\frac{\pr}{\pr x}\Delta t \eta \cos\omega t + D\Delta t\frac{\pr^2}{\pr x^2}\right]\nonumber\\
&&\times\delta(x-x^\p)\nonumber\\
&& =\frac{1}{2\pi}\int_{-\infty}^{+\infty} d\lambda\exp[\frac{\pr}{\pr x}\Delta t \eta \cos\omega t + D\Delta t\frac{\pr^2}{\pr x^2}]e^{i\lambda (x-x^\p)}\nonumber\\
&&= \frac{1}{2\pi}\int_{-\infty}^{+\infty} d\lambda e^{[-D \lambda^2 \Delta t + i\Delta  t\lambda (\frac{x-x^\p}{\Delta t} +\eta\cos\omega t)]}.
\end{eqnarray}
Standard Gaussian integral for $\lambda$ gives
\begin{equation}
\Fa =\frac{1}{\sqrt{4\pi D \Delta t}}e^{-\frac{1}{4D}(\frac{x-x^\p}{\Delta t}+\eta\cos\omega t)^2\Delta t}.
\end{equation}
Using Chapman-Kolmogorov equation the transition probability for a finite time interval is given by
\begin{eqnarray}
&&\Fb = \nonumber\\
&&\int_{x_0}^{x_t}Dx_s\exp{\left[-\frac{1}{4D}\int_{t_0}^t ds (\dot{x}_s +\eta\cos\omega s)^2 \right]}\nonumber\\
&&=\int_{x_0}^{x_t}Dx_s\exp{\left[\int_{t_0}^t ds L(\dot{x}_s) \right]}.
\end{eqnarray}
Here, $L(\dot{x}_s)=-\frac{1}{4D}(\dot{x}_s +\eta\cos\omega s)^2$ is the Lagrangian for this driven Brownian motion.
%%%%%%%%%%%%%%%%%%%%%%%%%%%%%%%%%%%%%%%%%%%
%\section*{Appendix C:}


\begin{thebibliography}{100}
\bibitem{com1} It is to be noted that in other treatments the B-particle is trapped in a harmonic potential and the potential is pulled at constant speed.
\end{thebibliography}
\end{document}